\documentclass[aps,prl,twocolumn,groupedaddress]{revtex4}
\usepackage{graphicx,epsfig}
\usepackage{subeqn}
\date{\today}

\newcommand{\cZ}{{\mathcal Z}}

\begin{document}
\title{Comment on ``Statistical Distribution for Generalized Ideal Gas of Fractional-Statistics Particles'', Phys. Rev. Lett. {\bf 73}, 922 (1994)}
\author{Drago\c s-Victor Anghel}
\affiliation{Department of Theoretical Physics, National Institute for
  Physics and Nuclear Engineering--''Horia Hulubei'', Str. Atomi\c stilor
  no.407, P.O.BOX MG-6, M\u agurele, Jud. Ilfov, Romania}

\maketitle

In Ref. \cite{PhysRevLett.67.937.1991.Haldane} Haldane introduced 
the fruitful concept of 
\textit{fractional exclusion statistics} (FES). One of the most influential 
papers in which the thermodynamics of FES systems was deduced is 
Ref. \cite{PhysRevLett.73.922.1994.Wu}. Unfortunately, some important, but 
eventually subtle, properties of the exclusion statistics parameters were 
overlooked in the original paper \cite{PhysRevLett.67.937.1991.Haldane} 
and in all the papers after that, including Ref. 
\cite{PhysRevLett.73.922.1994.Wu}. 
This omission makes the thermodynamics of FES systems inconsistent 
when mutual exclusion statistics is manifesting in the system 
(see Ref. \cite{JPhysA.40.F1013.2007.Anghel} for details). 
By this Comment I want to point-out this error--an error that persisted 
for such a long time--and to  give the correct statistical mechanics 
interpretation of FES. 

%draw attention of the author of Ref. \cite{PhysRevLett.73.922.1994.Wu} (and of the readership in general) of the error that exists in his paper--and persisted for such a long time--and to give the correct results (see Eqs. \ref{Wu_eq_pop} vs. Eq. \ref{inteq_for_n1}). I do this by giving in the same time a very simple and intuitive proof of a conjecture introduced in Ref. \cite{JPhysA.40.F1013.2007.Anghel}. 

For brevity, I shall use the notations and definitions of Ref. 
\cite{PhysRevLett.73.922.1994.Wu}, mainly without explanations.  
Let us assume that we have a FES system containing the species 
of particles indexed by $i=0,1,\ldots$. 
%and we denote by $\alpha_{ij}$ the exclusion statistics parameters. 
In the formalism of Ref. \cite{PhysRevLett.73.922.1994.Wu}, the number of 
configurations and the entropy of the 
system are given by (see Eqs. 4 and 6 of \cite{PhysRevLett.73.922.1994.Wu}) 
%
%\begin{equation} \label{conf_number1}
$W = \prod_i\frac{\left[G_i+N_i-1-\sum_j\alpha_{ij}(N_j-\delta_{ij})
\right]!}
{N_i!\left[G_i-1-\sum_j\alpha_{ij}(N_j-\delta_{ij})\right]!}$
%\end{equation}
%
and $\cZ=W \prod_i e^{\beta(\mu_i-\epsilon_i)}$, respectively, where $G_i$ and 
$N_i$ are the number of states and the number of particles of species $i$. 
Maximizing $\cZ$ 
%using the procedure of Ref. \cite{PhysRevLett.73.922.1994.Wu}, we 
one gets the system of equations 
\begin{subequations}\label{Wu_eq_pop}
\begin{eqnarray}
&& (1+w_i)\prod_j\left(\frac{w_j}{1+w_j}\right)^{\alpha_{ji}} = 
e^{\beta(\epsilon_i-\mu_i)} \label{EqforwWu} \\
%\end{equation}
%
%where $n_i$s can be calculated from the new system, 
%
%\begin{equation}
&& \sum_j(\delta_{ij}w_j+\beta_{ij})n_j = 1 , \label{system_ni_Wu}
\end{eqnarray}
\end{subequations}
where $n_i=N_i/G_i$ and $\beta_{ij}=\alpha_{ij}G_j/G_i$. 
As shown in \cite{JPhysA.40.F1013.2007.Anghel}, if there are 
$\alpha_{ij}\ne 0$ for $i\ne j$, 
then the system (\ref{Wu_eq_pop}) would be physically inconsistent 
since the populations $n_i$ would depend on the decomposition of the 
original system into the subsystems $i=0,1,\ldots$, which is in general 
arbitrary to a large extent--this being the standard coarse-graining 
procedure. In the same paper I corrected this 
inconsistency by a conjecture: I replaced $\alpha_{ij}$ by 
$\tilde\alpha_{ij}$, which, for $i\ne j$ are proportional to the 
dimension of the space on which they act; concretely, 
\begin{equation}
\tilde\alpha_{ii}=c_{ii}\quad {\rm and}\quad \tilde\alpha_{ij}=G_i c_{ij},
\quad {\rm for}\quad i\ne j. \label{extensive}
\end{equation}
%where $c_{ij}$ are constants. 
With these new statistics parameters, the system of equations,
\begin{equation}
\beta(\mu_i-\epsilon_i)+\ln\frac{[1+\tilde n_i]^{1-c_{ii}}}
{\tilde n_i} = \sum_{j(\ne i)} G_{j}\ln[1+\tilde n_{j}] 
c_{ji}, \label{inteq_for_n1}
\end{equation}
for the single-particle level populations replaces Eqs. (\ref{Wu_eq_pop}), 
which are the same as Eqs. (10) and (11) in Ref. 
\cite{PhysRevLett.73.922.1994.Wu}. 

In Ref. \cite{PhysLettA.372.5745.2008.Anghel} I showed that 
FES is manifesting in general interacting systems and I calculated the 
exclusion statistics parameters. In this way I showed that the mutual 
exclusion statistics parameters satisfy indeed the conjecture introduced 
in \cite{JPhysA.40.F1013.2007.Anghel}. 

Finally, in Ref. \cite{arXiv:0906.4836.Anghel} I analysed the basic properties of 
the exclusion statistics parameters using a vey simple and intuitive model. 
In this model I assumed, like above, that we have a system consisting of 
the particle species $i=0,1,\ldots$. If we split one of the species, say 
species $j$, into a number of sub-species, 
$j_0,j_1,\ldots$, then all the parameters $\tilde\alpha_{kl}$, with 
both, $k$ and $l$ different from $j$, remain unchanged, whereas the 
rest of the parameters must satisfy the relations
\begin{subequations}\label{relalphasgen}
\begin{eqnarray}
\tilde\alpha_{ij} &=& \tilde\alpha_{ij_0}=\tilde\alpha_{ij_1}=\ldots,\ 
{\rm for\ any}\ i,\ i\ne j \label{alphaij} \\
\tilde\alpha_{ji} &=& \tilde\alpha_{j_0i}+\tilde\alpha_{j_1i}+\ldots,\ 
{\rm for\ any}\ i,\ i\ne j \label{alphaji} \\
\tilde\alpha_{jj} &=& \tilde\alpha_{j_0j_0}+\tilde\alpha_{j_1j_0}+\ldots 
\nonumber \\
&=& \tilde\alpha_{j_0j_1}+\tilde\alpha_{j_1j_1}+\ldots = \ldots
\label{alphajj}
\end{eqnarray}
\end{subequations}
\textit{These are the general properties that have to be satisfied by the 
exclusion statistics parameters in any system of FES.}

Notice now that the conjecture (\ref{extensive}) introduced in Ref. 
\cite{JPhysA.40.F1013.2007.Anghel} satisfies 
Eqs. (\ref{relalphasgen}). Eventually in any physical system one can 
do the coarse graining finely enough, so that the exclusion statistics 
parameters satisfy Eq. (\ref{extensive}), as happened in the systems 
analysed in Ref. \cite{PhysLettA.372.5745.2008.Anghel}. In such a 
case the  particle populations of the energy levels are given by 
Eqs. (\ref{inteq_for_n1}). 

Equations (\ref{Wu_eq_pop}) (or 10 and 11 in Ref. 
\cite{PhysRevLett.73.922.1994.Wu}) are valid only if 
$\tilde\alpha_{ij}=0$ for any $i\ne j$. In such a case 
the result is identical to the result of (\ref{inteq_for_n1}). 

%\bibliography{/home/dragos/general}

\end{document}